\documentstyle[multicol,aps,prl,epsf]{revtex}
\begin{document}
\title{Polymer collapse in the presence of
hydrodynamic interactions}
\author{N. Kikuchi, A. Gent and J. M. Yeomans}
\address{Department of Physics, Theoretical Physics, 1 Keble Road,
Oxford, OX1 3NP, England.
}
\date{\today}
\maketitle
\begin{abstract}
We investigate numerically the dynamical behaviour of a polymer chain 
collapsing in a dilute solution. The rate of collapse is measured with 
and without the presence of hydrodynamic interactions. We find that 
hydrodynamic interactions both accelerate polymer collapse and alter 
the folding pathway.
\end{abstract}

\begin{multicols}{2}
\section{Introduction}

When a polymer is placed in solution hydrophobic interactions 
between monomers and solvent molecules can cause it to undergo a 
collapse transition to a compact state\cite{DG79,DE86}. The statistical 
physics of the polymer transition from the extended to the collapsed 
state is well understood. However the dynamics of the transition 
and in particular the effect on that dynamics of the hydrodynamic 
properties of the solvent remains unclear. Therefore in this Letter 
we investigate numerically the dynamical behaviour of a polymer chain 
collapsing in a dilute solution. The collapse is measured with and 
without the presence of hydrodynamic interactions thus allowing a 
direct investigation of their effect. We find that hydrodynamics 
accelerates the polymer collapse. It also alters the folding pathway, 
allowing the folding to occur more homogeneously along the polymer 
chain rather than initially at the chain ends.

With the introduction of Zimm's model\cite{DE86} it became 
apparent that hydrodynamics play a central role in the dynamics 
of polymers in dilute solution. However, understanding such 
interactions is difficult, analytically because they present a 
complicated many-body problem, and numerically because they 
develop on time-scales long compared to the thermal fluctuations 
of the monomers.

Theoretical work on the dynamics of polymer 
collapse can be divided into two approaches. Phenomenological 
models balance the driving and dissipative forces to give scaling 
laws\cite{DG85,OB94,BB96,HG00,AL01}. 
They involve assumptions about how the collapsed state 
develops on which there is no consensus. Several authors have considered models which are based on a
solution of the Langevin equation\cite{TD95,PO98,GL95}. Of particular interest
is work by Pitard\cite{P99} and by Kuznetsov et al\cite{KT96} who find the
inclusion of hydrodynamics, modelled by a  preaveraged Oseen tensor, speeds
up the collapse.

Simulations on polymer collapse\cite{OB94,BK95,KT95,SM00}, using Monte
Carlo or Langevin approaches, have not included the hydrodynamic effects of
the solvent. Very recent work has shown that it is now possible to use
molecular dynamics simulations with an explicit solvent to model the
collapse transition if powerful computational resources are available\cite{AL01}. In an interesting recent paper Chang and Yethiraj\cite{CY01} compared
molecular dynamics simulations of a polymer in a solvent to Brownian
dynamics simulations. They attempted to match the parameters in the two
simulations and hence to compare collapse with and without hydrodynamics. Here we use a hybrid approach\cite{MY00} 
where the solvent is modelled by a Malevanets-Kapral 
method\cite{MK99} and the polymer by molecular dynamics 
in our investigation of the hydrodynamics of polymer collapse.

\section{Simulation details}

Modelling a dilute polymer solution is a difficult task because 
of the existence of widely differing time scales. The dynamical 
properties of polymers can be dominated by hydrodynamic interactions 
between different parts of the polymer chains\cite{DE86}. In contrast 
with the time scale of thermal fluctuations of individual monomers, 
these interactions are long-ranged and evolve slowly. Therefore it 
is computationally too expensive to reach hydrodynamic time scales 
using molecular dynamics simulations for both the polymer and the 
solvent molecules.

To overcome this problem we use a hybrid simulation approach 
where the equations of motion of the polymer alone are solved using 
a molecular dynamics algorithm. The solvent is modelled using a 
mesoscale approach, developed by Malevanets and Kapral\cite{MK99}. 
This ignores the molecular detail of the solvent but preserves its 
ability to transmit hydrodynamic forces. The polymer can be thought 
of as moving within a ``hydrodynamic heat bath''.

The polymer chain is modelled by beads connected via 
non-harmonic springs\cite{KG90} with adjacent beads along the 
chain backbone representing an effective Kuhn length of the polymer 
chain. These finitely extensible springs are represented by the 
FENE potential
\begin{eqnarray}
\mbox{$V_{FENE}$}\left(\mbox{$r$}\right) = -\frac{\kappa}{2} 
{R_0}^2\ln\left[1-{\left(\frac{r}{R_0}\right)}^2\right],\quad 
\mbox{$r$}\,\mbox{$<$}\, R_o. \label{FENE}
\end{eqnarray}
A Lennard-Jones potential\cite{AT89} which acts between all the 
polymer beads is used to model the excluded volume of the monomers 
and a long range attraction which drives polymer collapse
\begin{eqnarray}
\mbox{$V_{LJ}$}\!\left(\mbox{$r$}\right)=
4\varepsilon\!\left[ {\left(\frac{\sigma}{r}\right)}^{\!12}-
{\left(\frac{\sigma}{r}\right)}^{\!6}\,\right].\label{LJ}
\end{eqnarray}
We take \(\varepsilon=1.0\), \(\sigma=1.0,\,\, \kappa=30\) and 
\({R_0}=2\) where parameters and results are quoted in reduced 
Lennard-Jones units.

Newton's equations of motion for the polymer are integrated 
using the time reversible velocity Verlet algorithm\cite{AT89}. 
The molecular dynamics time step is chosen to be
\({\delta}t=0.002\,t_s\) 
where \(t_s\) is the interval between solvent collision steps, 
defined below.

The solvent is modelled by a large number \(N=131072\) of 
point-like particles which move in continuous space with continuous 
velocities but discretely in time \cite{MK99}. The algorithm is 
separated into two stages. In the first of these, a free streaming 
step, the positions of the solvent particles at time \(t\), 
\({\mbox{\boldmath $x$}}_i(t)\), are updated simultaneously according to
\begin{eqnarray}
{\mbox{\boldmath $x$}}_i\,(t\!+\!t_s)\,=\,{\mbox{\boldmath $x$}}_i\,
(t)+{\mbox{\boldmath $v$}}_i\,(t)t_s
\end{eqnarray}
where \({\mbox{\boldmath $v$}}_i(t)\) is the velocity of a particle.

The second component of the algorithm is a collision step 
which is executed on both solvent particles and polymer beads. 
The system is coarse-grained into \(L^3\) unit cells of a regular 
cubic lattice. In this simulation \(L=32\) is used. There is no 
restriction on the total number of solvent or polymer particles 
in each cell, although the total number of particles is 
conserved. Multiparticle collisions are performed within each 
individual cell of the coarse-grained system by rotating the 
velocity of each particle relative to the centre of mass 
velocity \({\mbox{\boldmath $v$}}_{cm}(t)\) of all the particles 
within that cell
\begin{eqnarray}
{\mbox{\boldmath $v$}}_i\,(t\!+\!t_s)\,=\,{\mbox{\boldmath $v$}}_{cm}
(t)+{\mbox{\boldmath $R$}}\left(\,{\mbox{\boldmath $v$}}_i\,(t)-
{\mbox{\boldmath $v$}}_{cm}(t)\,\right).
\end{eqnarray}
\({\mbox{\boldmath $R$}}\) is a rotation matrix which rotates 
velocities by \(\theta\) around an axis generated randomly for 
each cell and at each time step. In the present calculations 
we take \(\theta=\frac{\pi}{2}\).

Note that the collision step preserves the position of the 
solvent and polymer beads. It transfers momentum between the 
particles within a given cell while conserving the total 
momentum of these particles. Because both momentum and energy 
are conserved locally the thermohydrodynamic equations of motion 
are captured in the continuum limit\cite{MK99}. Hence hydrodynamic 
interactions can be propagated by the solvent and, because the 
polymer beads are involved in the collisions, to the polymer. Note, 
however, that molecular details of the solvent are excluded: this 
allows the hydrodynamic interactions to be modelled with minimal 
computational expense.

The volume in phase space is invariant under both the free 
streaming and collision steps. Hence the system is described by 
a microcanonical distribution at equilibrium\cite{MK99}. The 
initial solvent distribution was generated by assigning positions 
randomly within the system with an average density \(\rho=4\) 
particles per unit cell. The velocities were assigned from a 
uniform distribution which relaxed rapidly (\(t_s\leq100\)) to 
the equilibrium Maxwell-Boltzmann form.

A particularly useful feature of the Malevanets-Kapral 
algorithm is the ease with which hydrodynamic interactions can 
be ``turned-off'' thus replacing the hydrodynamic heat bath by a 
Brownian (random) heat bath. This is achieved by randomly 
interchanging the velocities of all the solvent particles after 
each collision step, thus relaxing the constraint of local 
momentum conservation to a global one. Accordingly the velocity 
correlations which result in hydrodynamic interactions disappear 
from the fluid. Running simulations with the same initial conditions 
and parameter values, but with hydrodynamics present or absent, 
greatly faciliates pinpointing the effect of the hydrodynamic interactions.

\section{Results}

Our aim is to study the dynamics of polymer collapse from an 
extended to a compact state\cite{DG79}. The collapse 
trasnsition is driven by the attractive Lennard-Jones interactions 
between the polymer beads. The transition is rounded 
and shifted by the finite chain length, and takes place at 
\(k_BT\sim1.8\) for chains of length \(N_p=100\).

Initial polymer configurations were chosen at random from 
long runs on a  polymer chain at equilibrium in a solvent with 
\(k_BT=4\). Each of these extended configurations were placed 
in a solvent at equilibrium at \(k_BT=0.8\). This value ensures 
collapse but prevents the chain collapsing so quickly that 
hydrodynamic interactions do not have sufficient time to develop. 
The rate of polymer collapse was measured by monitoring the 
variation of the radius of gyration \(R_g\) of the chain with time
\begin{eqnarray}
\mbox{${R_g}^2$}(t)&=&\frac{1}{N_p}\sum_{i=1}^{N_p}{\left(R_i(t)
-R_{cm}(t)\right)}^2, \label{Rg} \nonumber\\
\mbox{$R_{cm}(t)$}&=&\frac{1}{N_p}\sum_{i=1}^{N_p}R_i(t). 
\label{Rcm} 
\end{eqnarray}
A typical numerical result is shown in Figure 1 for a chain of 
length \(N_p=100\) with and without hydrodynamics.

The collapse time was estimated as the time when the 
equilibrium radius of gyration was first attained. Results 
for \(\tau\) are shown for chains of varying lengths in Table 1. 
\(\tau\) was averaged over 20 initial configurations for 
\(N_{p}=40\), \(60\) and \(100\), and 5 for \(N_{p}=200\), and 
the collapse time for hydrodynamic ($\tau_H$) and Brownian ($\tau_B$) 
heat baths 
compared. Variations in collapse times between different runs are 
large as expected. However, it is strikingly apparent that 
hydrodynamic interactions speed up the rate of collapse by a 
factor\(\,\sim2\) for each polymer length. For all but 6 of the 
65 runs and for all the \(N_{p}=100\) and \(200\) runs the collapse 
was faster with hydrodynamics switched on. The results in Table 1 
also illustrate that the collapse time increases with chain length 
as expected ($\tau_B \sim N_p^{1.6 \pm 0.2}$, 
$\tau_H \sim N_p^{1.2 \pm 0.3}$). Values predicted in the literature for the exponents relating the collapse
time to chain length vary widely. It is far from obvious that these
exponents are universal: they appear to depend on quench depth and the
details of the model\cite{CY01}. Moreover the final radius of
gyration was recorded 
for both the Brownian and hydrodynamic collapses and found to 
agree as the equilibrium polymer properties should be independent 
of the nature of the heat bath.

Figure 2 compares typical collapse pathways with and 
without hydrodynamics for \(N_{p}=100\) and a final temperature 
\(k_BT=0.8\). In this and the majority of other runs a 
qualitative difference was observed in the collapse mechanism. 
In the Brownian solvent the ends of the polymer tend to collapse 
first forming a dumbbell shape. With hydrodynamics the collapse 
takes place more evenly along the chain. We speculate that this occurs due to 'slip-streaming'. The movement of beads
towards each other, which is initiated by the attractive Lennard-Jones
potential, is enhanced by hydrodynamic interactions. Once a bead starts
moving in a given direction and locally drags fluid with it, it is easier
for neighbouring beads to move in the same direction. Similar behaviour albeit on much 
larger length scales has been observed experimentally by Bartlett 
et al\cite{BH01} for a pair of colloidal particles in solution.
\section{Discussion}

We have shown that it is possible to measure the dynamics of 
the collapse transition of a polymer in a solvent using a hybrid 
mesoscale/molecular dynamics algorithm. Hence it has been possible 
to show that hydrodynamic interactions speed up the collapse of 
the polymer chain. The hydrodynamics alters the collapse pathway, 
allowing the folding to occur more homogeneously along the chain, 
rather than initially at the chain ends.

Qualitative features of the collapse pathways are in agreement with those
reported by Chang and Yethiraj\cite{CY01} who recently compared molecular dynamics
simulations of a polymer in a solvent, which include hydrodynamics, to
Brownian dynamics simulations, which did not. These authors also found that
the Brownian simulations could become trapped in a metastable free energy
minimum. It would be interesting to address whether this is due to deeper
quenches than those considered here or to the different simulation
approaches employed.

The solvent is modelled using the Malevanets-Kapral algorithm 
which sustains hydrodynamic modes but does not include molecular 
interactions. Hence we caution that we cannot investigate the late 
stages of collapse where trapping of the water molecules in 
the polymer chain may be important. Moreover collapse is driven 
by attractive Lennard-Jones interactions between the polymer beads 
rather than repulsive monomer-solvent interactions. It would be of 
interest to include the solvent near the polymer in the more 
realistic molecular dynamics updating and the possibility of doing 
this is under investigation.

Finally it may be of interest to consider whether the early 
stages of the folding of proteins in aqueous solution might 
be influenced by the hydrodynamic interactions of the solvent 
increasing the cooperativity of the collapse.

{\bf Acknowledgements}
We thank A. Malevanets, C. Pooley and P. Warren for helpful discussions.


\end{multicols}
\newpage

\begin{figure}[hbp]
\begin{center}
\centerline{\epsfxsize=12cm \epsffile{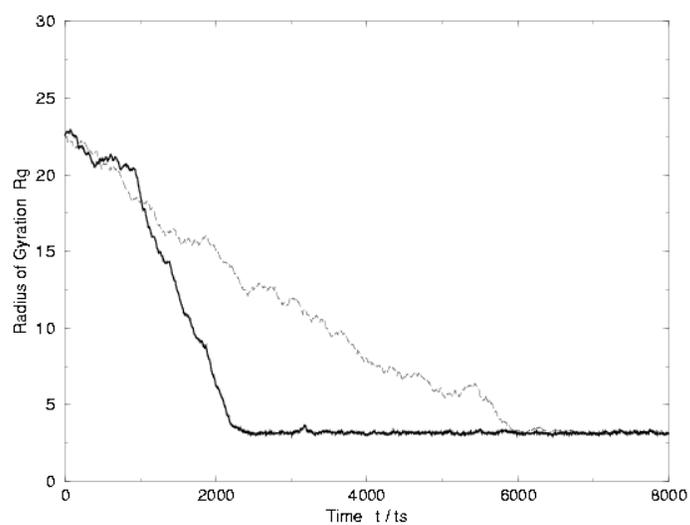} }
\end{center}
\caption{Variation of the radius of gyration with time for a 
collapsing polymer chain in a hydrodynamic (---) or Brownian (\(\cdots\)) heat bath.}
\end{figure}

\begin{figure}[hbp]
\begin{center}
\centerline{\epsfxsize=12cm \epsffile{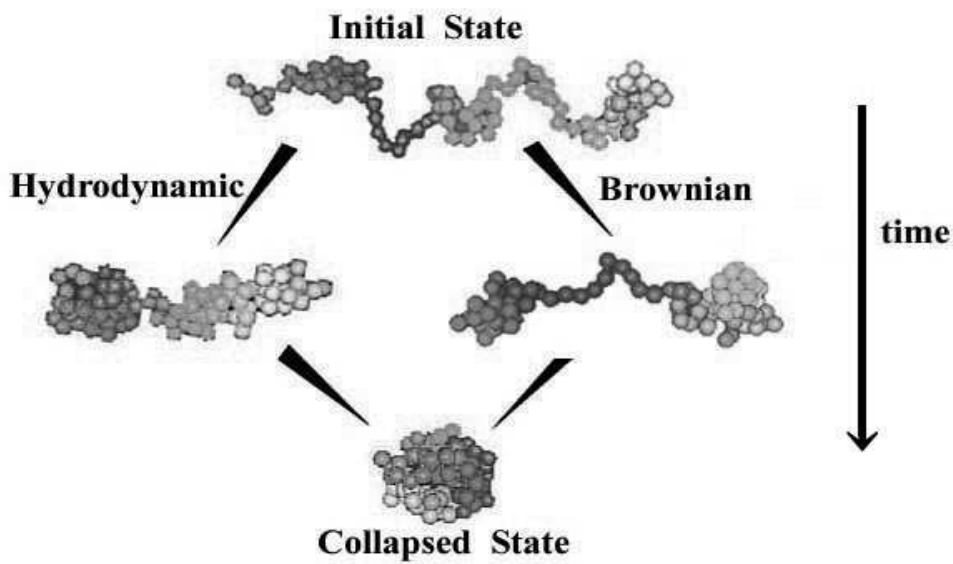} }
\end{center}
\caption{A comparison of the pathways for polymer collapse with and without hydrodynamics.}
\end{figure}

\newpage
\begin{table}[hbp]
\label{collapsetime}
\begin{center}
\begin{tabular}{|c|c|c|c|c|}\hline
\(N_p\) & \(\left<{\tau}_H\right>\) & \(\left<{\tau}_{B}\right>\) & \(\left<{R_g}_H\right>\) & \(\left<{R_g}_{B}\right>\) \\ \hline
40 & 290 $\pm$ 92 & 401 $\pm$ 118 & 2.02 $\pm$ 0.05 & 2.02 $\pm$ 0.04 \\ \hline
60 & 490 $\pm$ 79 & 671 $\pm$ 170 & 2.23 $\pm$ 0.02 & 2.23 $\pm$ 0.01 \\ \hline
100 & 858 $\pm$ 113 & 1571 $\pm$ 252 & 2.58 $\pm$ 0.01 & 2.58 $\pm$ 0.02 \\ \hline
200 & 2050 $\pm$ 253 & 5500 $\pm$ 390 & 3.15 $\pm$ 0.00 & 3.16 $\pm$ 0.02 \\ \hline
\end{tabular}
\end{center}
\caption{Averaged collapse time in units of the solvent time step
\(t_s\) of a polymer chain of length \(N_p\) with
(\(\left<{\tau}_H\right>\)) and without (\(\left<{\tau}_B\right>\))
hydrodynamics for a final temperature \(k_BT=0.8\).
       The radius of gyration after collapse in reduced Lennard-Jones units with (\(\left<{R_g}_H\right>\)) and without (\(\left<{R_g}_B\right>\)) hydrodynamics is also listed.}
\end{table}

\end{document}